\PassOptionsToPackage{unicode}{hyperref}
\PassOptionsToPackage{hyphens}{url}
\documentclass[
  11pt,
]{article}
\usepackage{lmodern}
\usepackage{amssymb,amsmath}
\usepackage{ifxetex,ifluatex}
\ifnum 0\ifxetex 1\fi\ifluatex 1\fi=0 % if pdftex
  \usepackage[T1]{fontenc}
  \usepackage[utf8]{inputenc}
  \usepackage{textcomp} % provide euro and other symbols
\else % if luatex or xetex
  \usepackage{unicode-math}
  \defaultfontfeatures{Scale=MatchLowercase}
  \defaultfontfeatures[\rmfamily]{Ligatures=TeX,Scale=1}
\fi
\IfFileExists{upquote.sty}{\usepackage{upquote}}{}
\IfFileExists{microtype.sty}{% use microtype if available
  \usepackage[]{microtype}
  \UseMicrotypeSet[protrusion]{basicmath} % disable protrusion for tt fonts
}{}
\makeatletter
\@ifundefined{KOMAClassName}{% if non-KOMA class
  \IfFileExists{parskip.sty}{%
    \usepackage{parskip}
  }{% else
    \setlength{\parindent}{0pt}
    \setlength{\parskip}{6pt plus 2pt minus 1pt}
    }
}{% if KOMA class
  \KOMAoptions{parskip=half}}
\makeatother
\usepackage{xcolor}
\IfFileExists{xurl.sty}{\usepackage{xurl}}{} % add URL line breaks if available
\usepackage[margin=2.5cm]{geometry}
\usepackage{graphicx}
\makeatletter
\def\maxwidth{\ifdim\Gin@nat@width>\linewidth\linewidth\else\Gin@nat@width\fi}
\def\maxheight{\ifdim\Gin@nat@height>\textheight\textheight\else\Gin@nat@height\fi}
\makeatother
\setkeys{Gin}{width=\maxwidth,height=\maxheight,keepaspectratio}
\makeatletter
\def\fps@figure{htbp}
\makeatother
\ifluatex
  \usepackage{selnolig}  % disable illegal ligatures
\fi
\usepackage[style=numeric,]{biblatex}
\title{The effects of H blistering and ELMs on the thermal fatigue
cracking of W by strikepoint sweeping}
\author{true \and true \and true \and true}
\date{}

\usepackage{kantlipsum}

\usepackage{abstract}
\makeatletter
\def\@maketitle{%
  \newpage
  \let \footnote \thanks
      {\fontsize{20}{20}\selectfont\raggedright  \setlength{\parindent}{0pt} \@title \par}
    }
\makeatother

\title{The effects of H blistering and ELMs on the thermal fatigue
cracking of W by strikepoint sweeping }

\date{}

\usepackage{titlesec}

\titleformat*{\section}{\large\bfseries}
\titleformat*{\subsection}{\normalsize\itshape} % \small\uppercase
\titleformat*{\subsubsection}{\normalsize\itshape}
\titleformat*{\paragraph}{\normalsize\itshape}
\titleformat*{\subparagraph}{\normalsize\itshape}

\usepackage[section]{placeins}

\makeatletter
\@ifpackageloaded{hyperref}{}{%
\ifxetex
  \PassOptionsToPackage{hyphens}{url}\usepackage[setpagesize=false, % page size defined by xetex
              unicode=false, % unicode breaks when used with xetex
              xetex]{hyperref}
\else
  \PassOptionsToPackage{hyphens}{url}\usepackage[draft,unicode=true]{hyperref}
\fi
}

\@ifpackageloaded{color}{
    \PassOptionsToPackage{usenames,dvipsnames}{color}
}{%
    \usepackage[usenames,dvipsnames]{color}
}
\makeatother
\hypersetup{breaklinks=true,
            bookmarks=true,
            pdfauthor={J. Hargreaves, J. Vernimmen, J. Scholten, T.W
Morgan},
             pdfkeywords = {tungsten, fatigue, cracking, blistering,
divertor lifetime, plasma-material interactions},
            pdftitle={The effects of H blistering and ELMs on the
thermal fatigue cracking of W by strikepoint sweeping},
            colorlinks=true,
            citecolor=blue,
            urlcolor=blue,
            linkcolor=magenta,
            pdfborder={0 0 0}}
\makeatletter
\def\fps@figure{htbp}
\makeatother

\usepackage{caption}
\usepackage{floatrow}

\DeclareNewFloatType{chunk}{placement=H, fileext=chk, name=}
\makeatletter
\@addtoreset{chunk}{section}
\makeatother

\usepackage{tabularx}
	\newcommand{\raisedrule}[3][0em]{\leavevmode\leaders\hbox{\rule[#1]{1pt}{#2}}\hfill\kern0pt \;\;\small\textsc{#3}\;\; \leavevmode\leaders\hbox{\rule[#1]{1pt}{#2}}\hfill\kern0pt}

\usepackage{longtable}
\usepackage[fixed]{fontawesome5}

\begin{document}

% \maketitle

{% \usefont{T1}{pnc}{m}{n}
\setlength{\parindent}{0pt}
\thispagestyle{plain}
{%\fontsize{18}{20}\selectfont\raggedright
\maketitle  % title \par
\vskip 8.5pt
}

{J. Hargreaves\textsuperscript{a,1}, {J.
Vernimmen\textsuperscript{a,2}, {J. Scholten\textsuperscript{a,3},
{ and T.W Morgan\textsuperscript{a,4}} \vskip 24.5pt

 % this endif is for removing titles in (anonymous)

\vskip -14.5pt

{\footnotesize

\begingroup
\setlength\parskip{-.1em}
				\textsuperscript{a} \emph{Dutch Institute for Fundamental Energy Research, TU/e Science Park, De Zaale 20, 5612 AJ Eindhoven, The Netherlands}\par
				\textsuperscript{b} \emph{Max Planck Institute for Plasma Physics, Boltzmannstraße 2, Garching, 85748, Germany}\par
				\textsuperscript{c} \emph{Eurofusion Consortium, Boltzmannstraße 2, Garching, 85748, Germany}
\endgroup

}

 % this endif is for removing affils in (anonymous)

% \vskip 8.5pt
%
% {\footnotesize
%
% \begingroup
% 	\noindent \emph{Email addresses:} % 		\textsuperscript{1} {\tt \href{mailto:j.p.hargreaves@differ.nl}{\nolinkurl{j.p.hargreaves@differ.nl}}},
% 		% 		\textsuperscript{2} {\tt \href{mailto:j.w.m.vernimmen@differ.nl}{\nolinkurl{j.w.m.vernimmen@differ.nl}}},
% 		% 		\textsuperscript{3} {\tt \href{mailto:j.scholten@differ.nl}{\nolinkurl{j.scholten@differ.nl}}},
% 		% 		\textsuperscript{4} {\tt \href{mailto:t.w.morgan@differ.nl}{\nolinkurl{t.w.morgan@differ.nl}}}%
% \endgroup
% }

 % if it's not anonymous, but it's still alternate layout

\begin{tabularx}{\linewidth}{@{}X@{}}
\begin{minipage}[t]{0.6\textwidth}

\raisedrule[0.2em]{0.1pt}{\footnotesize Abstract}

\vspace{5pt}

\footnotesize{Cyclic thermal loads imposed on EU DEMO's W divertor by
strikepoint sweeping may induce low-cycle thermal fatigue cracking of
its plasma-facing surfaces. This cracking may be accelerated by
plasma-material interactions such as H implantation, blistering, fuzz
and void formation. Fatigue cracking may also synergise with ELM
cracking. To explore these novel forms of environmentally-assisted
fatigue, FEA modelling was used to design a uniaxial fatigue experiment
for Magnum-PSI that represents strikepoint sweeping at 1 Hz across a 100
mm span of DEMO's divertor targets. Magnum-PSI was used to combine
cyclic thermal loading (850-1250°C) of W with H implantation (fluence
\textasciitilde10\textsuperscript{26} m\textsuperscript{-2}) and two
forms of ELM-like pre-cracking. Quantitative SEM analysis of
fatigue-cracked W revealed that H implantation significantly delayed
crack initiation, with pre-implanted targets requiring 450-600 cycles
before failure compared to \textless150 cycles for non-implanted
samples. This was attributed to hydrogen-induced dislocation pinning,
which produces a case-hardening effect that inhibits persistent slip
band formation. ELM-like pre-cracking combined with strikepoint sweeping
was found to give rise to localised melting and the formation of 30 µm
diameter droplets, caused by thermal isolation of W regions by fatigue
cracks. The implications for the fatigue lifetime of DEMO's divertor are
also discussed.}

\end{minipage}
		\hfill
		\begin{minipage}[t]{0.35\textwidth}

		\raisedrule[0.2em]{0.1pt}{\footnotesize Contact Info}

		\vspace{8pt}

		{\footnotesize

				\noindent 		\textsuperscript{1} {\tt \href{mailto:j.p.hargreaves@differ.nl}{\nolinkurl{j.p.hargreaves@differ.nl}}}\par
				\textsuperscript{2} {\tt \href{mailto:j.w.m.vernimmen@differ.nl}{\nolinkurl{j.w.m.vernimmen@differ.nl}}}\par
				\textsuperscript{3} {\tt \href{mailto:j.scholten@differ.nl}{\nolinkurl{j.scholten@differ.nl}}}\par
				\textsuperscript{4} {\tt \href{mailto:t.w.morgan@differ.nl}{\nolinkurl{t.w.morgan@differ.nl}}}				}
		\vskip 8.5pt

		\raisedrule[0.2em]{0.1pt}{\footnotesize Paper Info}

		\vspace{5pt}

			{\footnotesize

			\emph{Last updated}: 2 December 2025\vspace{1.5pt}\par
			\emph{Word Count}: 6,209

			 \par
			
			}

			\vskip 8.5pt

		\end{minipage}
%\bottomrule
%\rule[#1]{1pt}{#2}
\\
\\
\hline
\end{tabularx}

 % this endif is for futzing with the abstract in anonymous mode

 % for alternate-layout

\vskip -8.5pt

 % removetitleabstract

% We'll put doublespacing here
% Remember to cut it out later before bib
\section{1. Introduction}\label{introduction}

Ensuring divertor survival in a commercial fusion power plant will
require careful management of power and particle loads. EU DEMO's lower
single-null divertor will normally operate in a detached regime to keep
the thermal load on its tungsten (W) monoblock targets below their
technological limit of 10 MW m\textsuperscript{-2}
\autocite{siccinioFigureMeritDivertor2019,youEuropeanDEMODivertor2016}.
However, if the detached condition is lost (e.g.~due to impurity seeding
fault) the incident load is expected to increase to 45 MW
m\textsuperscript{-2} over 10 s
\autocite{mavigliaImpactPlasmaWallInteraction2021,youDiscussionReattachmentThermal2025}.
These slow thermal transients are predicted to exceed the divertor's
critical heat flux, resulting in severe deformation and melting of
plasma-facing surfaces
\autocite{mavigliaLimitationsTransientPower2016,youHighHeatFluxPerformanceLimit2022}.
To mitigate this it is proposed to sweep the plasma strikepoints at 1-5
Hz along a 50-400 mm span of the divertor targets during reattachment
events
\autocite{mavigliaLimitationsTransientPower2016,silburnMitigationDivertorHeat2017}.
This strikepoint sweeping would re-distribute the reattached thermal
load over a larger area, reducing monoblock surface temperatures and
ensuring divertor survival until the detached condition is retrieved.
However, the cylic thermal loads imposed on DEMO's W monoblocks by
strikepoint sweeping may give rise to plastic strain accumulation and
eventually result in low-cycle fatigue (LCF) cracking of the
plasma-facing surface
\autocite{mavigliaLimitationsTransientPower2016,liSweepingHeatFlux2016}.

During normal operation DEMO's monoblocks will be bombarded by intense
fluxes (10\textsuperscript{20} - 10\textsuperscript{24}
m\textsuperscript{-2} s\textsuperscript{-1}) of deuterium (D), tritium
(T) and helium (He) particles at 1-5 eV
\autocite{brezinsekPlasmaWallInteraction2017}. These energetic particle
loads will give rise to a range of plasma-material interactions,
including D/T/He implantation, void and blister formation, and the
formation of nano-scale W fuzz
\autocite{brezinsekPlasmaWallInteraction2017}. W sputtered by impurity
species may also re-deposit on plasma-facing surfaces, and exposure to
high temperatures may promote grain growth, recrystallisation, and creep
\autocite{wangSimulationTungstenTarget2021,pintsukLongPulseHighHeat2024}.
This evolution of surface morphology and local microstructure during
service may synergise with fatigue, altering the dislocation-mediated
initiation and microstructure-dependent early propagation of LCF cracks.

Little prior work exists on this topic, however, some inferences can be
made from literature on the thermal shock cracking of W by edge
localised modes (ELMs). One previous study used the Pilot-PSI linear
plasma device to expose W at 400 °C to a H fluence of 4 ×
10\textsuperscript{23} m\textsuperscript{-2}, followed by ELM-like
loading via Nd:YAG laser (1 ms at 0.64 GW m\textsuperscript{-2})
\autocite{wirtzImpactCombinedHydrogen2015}. W was found to be more prone
to cracking when combined with H implantation, which was attributed to H
embrittlement. These findings have recently been corroborated by an
experiment at the OLMAT neutral beam facility
\autocite{alegreFirstThermalFatigue2024}. A possible mechanism for this
is hydrogen-induced dislocation pinning (HIDP), which was observed via
nanoindentation and transmission electron microanalysis of
recrystallised W exposed at 50 °C to a D fluence of 1.2 ×
10\textsuperscript{24} m\textsuperscript{-2}
\autocite{liThreeMechanismsHydrogenInduced2020}. However, at higher
temperatures, H embrittlement is theorised to occur via hydrogen
enhanced local plasticity (HELP), which posits that the trapping of
implanted H at dislocation cores enhances dislocation mobility,
resulting in expediting pile-up and pinning by grain boundaries
\autocite{martinEnumerationHydrogenEnhancedLocalized2019}. These
mechanisms both affect slip and dislocation glide, and may therefore
alter persistent slip band (PSB) formation, crack initiation mechanics,
and the micro-scale propagation behaviour of LCF cracking
\autocite{ohIntegratedExperimentalComputational2024,sangidPhysicallyBasedFatigue2011}.

Cracking behaviour may also be altered by modification of plasma-facing
surface morphology
\autocite{ellyinFatigueDamageCrack2012,peguesSurfaceRoughnessEffects2018}.
An earlier Magnum-PSI study exposed W at 297 °C to a H plasma fluence of
10\textsuperscript{28} m\textsuperscript{-2} to create micron-scale
surface blisters and sub-surface voids, followed by 1 ms laser pulsing
with a \emph{ΔT} of 677-927 °C. Cracks were found to preferentially
initiate at blister edges due to the stress concentration effect, and a
greater number of smaller cracks were observed versus non-blistered
specimens \autocite{liInfluencePorosityBlistering2022}. These localising
effects may alter how micro-scale cracks coalesce into a dominant
fatigue crack, and voids may serve as microstructural inhomogeneities
through which cracks preferentially propagate
\autocite{ellyinFatigueDamageCrack2012,hanResearchEffectMicroVoids2023,brighentiInfluenceMaterialMicrovoids2014}.
Localised plastic deformation arising from void and blister formation
may also increase dislocation density, enhancing dislocation
entanglement
\autocite{sangidPhysicallyBasedFatigue2011,buziInfluenceTungstenMicrostructure2015}.

Understanding this novel form of environmentally-assisted fatigue will
be vital for lifetime analysis of DEMO's divertor
\autocite{wirtzImpactCombinedHydrogen2015,youStructuralLifetimeAssessment2021}.
This work addresses this via a campaign of novel uniaxial fatigue
experiments at Magnum-PSI, exploring the effects of H implantation and
prior ELM-cracking on the LCF cracking behaviour of W.
DEMO-representative experimental parameters are determined via
supporting finite element analyses (FEA), and cracking behaviour is
characterised ex-situ via a quantitative scanning electron microscopy
(SEM) study.

\section{2. Method}\label{method}

\subsection{2.1 Supporting finite element analyses
(FEA)}\label{supporting-finite-element-analyses-fea}

FEA was used to determine DEMO-representative thermomechanical loading
parameters for the Magnum-PSI experiment. This employed Siemens
Simcentre 3D 2306 with a time-dependent thermal-mechanical multiphysics
study. The step size was 0.01 s and the simulation time was 10.0 s.
Monoblock geometry was based on the DEMO baseline reduced width
ITER-like design with 8 mm armour
\autocite{youStructuralLifetimeAssessment2021},
\autocite{youHighHeatFluxTechnologiesEuropean2021}. Nonlinear
temperature-dependent relations for the thermal and mechanical
properties of W, Cu, and CuCrZr from the ITER material property handbook
were used, supplemented by a bilinear kinematic hardening model for W
\autocite{zinovevModelingStrainHardening2021}. Symmetry was exploited
such that only a quarter of the geometry was meshed. Initial conditions
for the transient analysis were determined by the steady state
simulation of a uniform 10 MW m\textsuperscript{-2} thermal load. The
load of the transient analysis assumed a Gaussian thermal flux footprint
for DEMO's strikepoints with a standard deviation (S.D,
\emph{σ\textsubscript{sp}}) of 21.2 mm, swept over the divertor targets
with a triangular waveform at a frequency (\emph{f\textsubscript{sw}})
of 1.0 Hz and a peak-to-peak amplitude (\emph{A\textsubscript{sw}}) of
100 mm
\autocite{liSweepingHeatFlux2016,ambrosinoSweepingControlPerformance2021}.
To simulate a gradual loss of divertor detachment the applied load also
increased linearly from 10 MW m\textsuperscript{-2}
(\emph{q\textsubscript{init}}) to 45 MW m\textsuperscript{-2}
(\emph{q\textsubscript{final}}) over 10 s
(\emph{t\textsubscript{reattach}})
\autocite{youDiscussionReattachmentThermal2025}. The resultant uniform
profile of heat flux incident on a central monoblock was approximated by
Eq. 1:

\begin{equation}
q(t) = \left(q_\mathrm{init} + \frac{\left(q\textsubscript{final}-q\textsubscript{init}\right) \times t}{t_\mathrm{reattach}}\right)\exp\left(-\frac{(A_\mathrm{sw}/2)^2\sin^2(2\pi f_\mathrm{sw}t)}{2\sigma_\mathrm{sp}^2}\right)
\end{equation}

Key parameters of the transient monoblock simulation are presented in
Fig. 1.

\begin{figure}
\centering
\includegraphics[width=1\textwidth,height=\textheight]{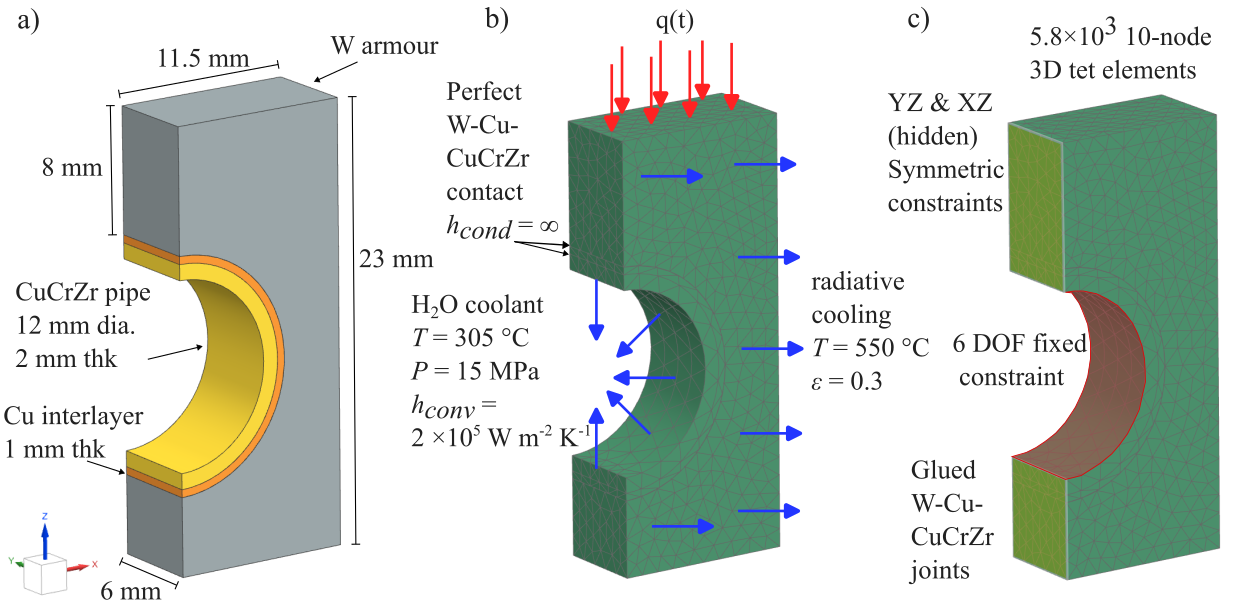}
\caption{Major dimensions (a), boundary conditions (b) and constraints
(c) of the monoblock FEA. Boundary conditions and parameters were taken
from
\autocite{edemettiOptimizationFirstWall2020,youDivertorEuropeanDEMO2022,el-morshedyThermalHydraulicModellingAnalysis2021}.
Neutronic heating has been neglected in the model.}
\end{figure}

A second FEA simulation was used to determine the thermal load required
for a Magnum-PSI target to match the thermomechanical response of a
monoblock undergoing strikepoint sweeping. This Magnum-PSI target
simulation employed the same study type, step size, and material
properties as the monoblock simulation, but model geometry was based on
a Magnum-PSI fatigue testing target (described in section 2.2). The
thermal load imposed by Magnum-PSI's beam was modelled as a radially
symmetric Gaussian with a S.D of 5.52 mm. Rotational symmetry was
exploited such that a quarter of the cylindrical target was modelled,
and the W and Cu domains were meshed into a total of
1.8×10\textsuperscript{4} 10-node parabolic tetrahedron (C3D10)
elements. Element size at the plasma-facing surface was set to 500 µm to
enhance local accuracy.

A glued joint with perfect thermal contact was applied to the brazed
W-Cu joint, and a fixed 21 °C temperature constraint was applied to the
target's base to represent ideal cooling of the target by Magnum-PSI's
water-cooled target holder. A fixed mechanical constraint (6DOF) was
applied to the same surface. Radiative cooling (\emph{ε} = 0.3) from
exposed surfaces to a 21 °C environment was assumed.

\subsection{2.2 Magnum-PSI target design \&
preparation}\label{magnum-psi-target-design-preparation}

Magnum-PSI fatigue testing targets consisted of a cylindrical puck
Ø25×12 mm of ITER-grade polycrystalline W (Plansee SE, Austria) brazed
to a oxygen-free high conductivity Cu disc Ø30×4 mm. W pucks were cut so
that plasma-facing surfaces were perpendicular to the rolling direction,
and Ø0.5 mm radial thermocouple hole was drilled into the side of each
OFHC Cu base to a depth of 15 mm.. At the centre of each plasma-facing
surface, a stress concentration notch 200 ± 20 µm in diameter was cut
via electrical discharge machining (EDM) to induce preferential crack
initiation at the target centre. Notch cutting damage was subsequently
removed by plane grinding using progressively finer SiC papers (\#180,
\#500, \#1000), followed by fine grinding using a Struer's MD-Allegro
composite diamond disc and 9 µm diamond paste. Final polishing employed
a 3 µm diamond solution and Struers MD-dur cloth followed by 0.25
colloidal silica and a Struers MD-chem cloth.

\subsection{2.3 H implantation and strikepoint sweeping exposures in
Magnum-PSI}\label{h-implantation-and-strikepoint-sweeping-exposures-in-magnum-psi}

The thermal effects of strikepoint sweeping was emulated by modulating
Magnum-PSI's cascaded arc plasma source with a 1 Hz sinewave, which
yielded a sinusoidal variation in target surface temperature
(\emph{T\textsubscript{surf}}). Plasma parameters (Table 1) were
monitored via Thomson scattering (TS), the laser of which (λ = 532 nm)
was operated at 10 Hz and synchronised with the plasma source current
modulation. This employed a perpendicular scattering geometry (θ = 90°)
with a chord length of 87 mm. H implantation plasma exposures were also
conducted. Further details on Magnum-PSI and its diagnostics can be
found in the literature
\autocite{vaneckHighfluenceHighfluxPerformance2019}.

\begin{table}[h!]
\caption{Plasma parameters for the H implantation and sweeping exposures. $\Gamma_{pk}$ denotes peak plasma flux at the plasma column centre, $\Phi$ denotes cumulative fluence, and DL denotes detection limit.}
\begin{tabular}{|c|c|c|c|c|c|c|c|}
\hline
Phase & Plasma & $T_e$ & $n_e$ \times 10\textsuperscript{21} & $\Gamma_{pk} \times 10\textsuperscript{23}$ & $T_{surf}$ & $t_{exp}$ & $\Phi$ \times 10\textsuperscript{26} \\
& Species & (eV) & (m\textsuperscript{-3}) &  (m\textsuperscript{-2} s\textsuperscript{-1}) & ($^{\circ}$C) & (s) & (m\textsuperscript{-2}) \\

\hline
H implantation & H (100\%) & 1.54 & 0.074 & 7.33 & 300 & 300 & 2.2 \\
1.0 Hz sweeping & H (100\%) & <DL - 1.6 & <DL - 1.5 & variable & 850 - 1250 & variable & variable \\
\hline
\end{tabular}
\end{table}

Target surface temperature was continuously monitored by a high
framerate infrared camera (FLIR SC7000MB) and a multi-wavelength
emissivity-independent pyrometer (FAR associates FMPI SpectroPyrometer).
This pyrometer was used to determine the emissivity (\emph{ε}) of the
polished target surface, which was found to be 0.10 - 0.17 between 600 -
1234 °C. Emissivity was re-measured at the start of every discharge and
the IR camera parameter adjusted (if necessary) to account for surface
roughening. Plasma composition was monitored via optical emission
spectroscopy (OES). To explore the combined effects of ELM-like
pre-cracking and strikepoint sweeping, three notchless targets with a
`technical' EDM-cut surface finish were also exposed. One notched and
polished target was also pre-cracked using a Nd:YaG laser (λ = 1064 nm)
prior to sweeping exposure. Laser parameters for this target are
summarised in Table 2.

\begin{table}[h!]
\caption{Laser parameters for simulating ELM cracking. Absorbed energy estimated based on an assumed polished surface emissivity of 0.1 and a beam transmission of 0.75.}
\begin{tabular}{|c|c|c|c|c|c|c|c|}
\hline
Pulse length & Frequency & No. pulses & Beam width & \multicolumn{2}{c|}{Per-pulse energy} & \multicolumn{2}{c|}{Surface temp.} \\
\cline{5-8}
(ms) & (Hz) & (-) & FWHM (mm) & emitted (J) & absorbed (J) & base ($^{\circ}$C) & delta ($^{\circ}$C) \\
\hline
1.0 & 10.0 & 10\textsuperscript{3} & 1.0 & 12.75 & 0.96 & 25 & 825 \\
\hline
\end{tabular}
\end{table}

\subsection{2.4 Ex-situ characterisation and quantitative image
analysis}\label{ex-situ-characterisation-and-quantitative-image-analysis}

Target surfaces were imaged after exposure using a ThermoFisher
Scientific Phenom Pharos field emission gun scanning electron microscope
(FEG-SEM). Secondary (SE) and backscattered (BSE) electron imaging
employed a beam voltage of 10 kV and a probe current of 7.5 nA. Energy
dispersive X-ray spectroscopy (EDX) was conducted at 20 kV. BSE quad
height reconstruction was used for surface profilometry. Prior to
analysis, each target was ultrasonically cleaned using deionised water +
detergent, acetone, ethanol, and isoproponol in sequence.

A 100 mm\textsuperscript{2} area of each target was imaged, centred upon
the notch. These high-resolution composite images were each comprised of
a 13 x 13 grid of 300x magnification BSE images, stitched together using
Fiji/ImageJ2's Microscopy Imaging Stitching Tool (MIST) plugin
\autocite{chalfounMISTAccurateScalable2017}. Stitching employed overlay
blending with a 5\% overlap, and stitched images were automatically
thresholded using Otsu's, Shabhang and triangle methods before
de-noising using a 3x3 median filter. The central notch and egregious
imaging defects (e.g.~surface debris) were masked prior to quantitative
image analysis. Total crack length and density were automatically
measured using Fiji's Ridge detection plugin, which implements Steger's
curvilinear structure detection algorithm
\autocite{stegerUnbiasedDetectorCurvilinear1998}.

\section{3. Results and discussion}\label{results-and-discussion}

\subsection{3.1 Supporting FEA}\label{supporting-fea}

Profiles of monoblock temperature, Von-Mises stress, and vertical (z)
displacement at the peak of the final sweeping cycle (\emph{t} = 9.87 s)
are shown in Fig. 2. The maximum temperature of the plasma-facing
surface was 3001 °C. At the centre of this surface (point A),
displacement perpendicular to the surface was found to be both at a
global maxima (\emph{ΔL\textsubscript{zz}} = 0.052 mm) and uniaxial
(i.e.~\emph{ΔL\textsubscript{xx}} = \emph{ΔL\textsubscript{yy}} = 0),
therefore the uniaxial true strain range (\emph{Δε\textsubscript{zz}})
at point A was found to be 0.06\% from the minimal and maximal vertical
displacement.

\begin{figure}
\centering
\includegraphics[width=1\textwidth,height=\textheight]{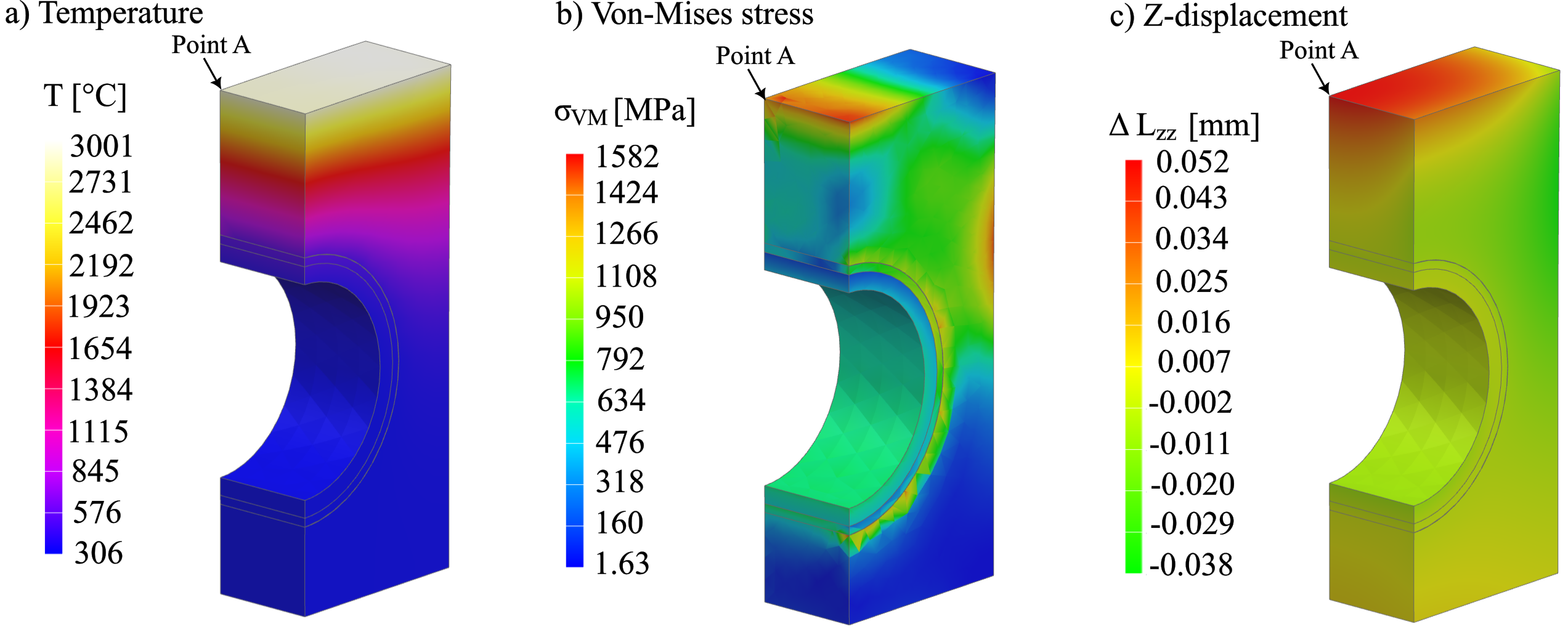}
\caption{Profiles of monoblock temperature (a), VM stress (b), and
z-displacement (c) profiles at \emph{t} = 9.87 s.}
\end{figure}

Similar profiles for the notched Magnum-PSI target at the peak of a
sweep are given in Fig. 3. The centrally located stress concentration
notch (point B) ensured that stresses at the plasma-facing surface
significantly exceeded tungsten's yield stress (346 MPa at 1200 °C) and
plastic deformation was achieved. Vertical displacement at point B was
also confirmed to be uniaxial and at a global maxima. The simulated
thermal load imposed by Magnum-PSI's Gaussian plasma column was
iteratively increased until the predicted uniaxial true strain range at
at the centre of the target matched the value calculated for the
monoblock. This was found to be achieved by a load of 20.25 MW
m\textsuperscript{-2} modulated at 1.0 Hz. As direct experimental
measurement of uniaxial target strain is not currently possible in
Magnum-PSI, the calculated surface temperature range of 842 - 1255 °C
was employed as an experimental parameter for the sweeping exposures.

\begin{figure}
\centering
\includegraphics[width=1\textwidth,height=\textheight]{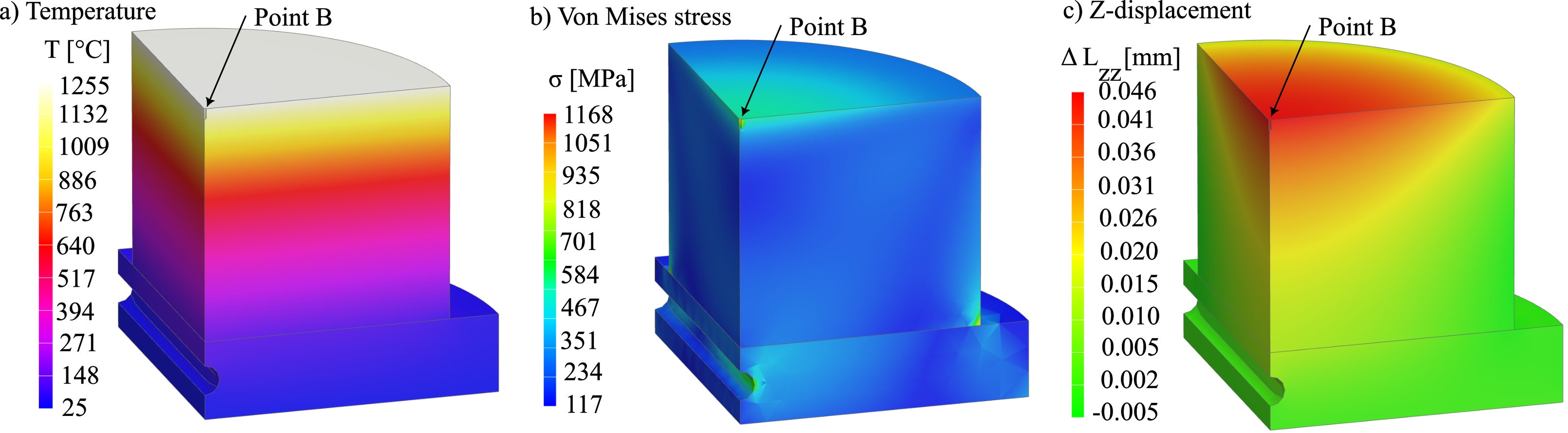}
\caption{Profiles of Magnum-PSI target temperature (a) VM stress (b) and
z-displacement at \emph{t} = 9.95s (c).}
\end{figure}

\newpage

Profiles of surface temperature (\emph{T\textsubscript{surf}}),
Von-Mises (VM) stress (\emph{σ\textsubscript{VM}}) and uniaxial true
strain range (\emph{Δε\textsubscript{zz}}) against time for Points A and
B are given by Fig. 4. While strikepoint sweeping is able to prevent
bulk W surface melting, temperatures in the upper portion of the
monoblock exceed 1350 °C approximately 2.6s after reattachment, thus
recrystallisation is highly likely to occur
\autocite{shahRecrystallizationBehaviourHighFlux2021}. The effects of
recrystalisation on cracking behaviour is presently intentionally
neglected by this experiment, but will be the subject of future work.

\begin{figure}
\centering
\includegraphics[width=1\textwidth,height=\textheight]{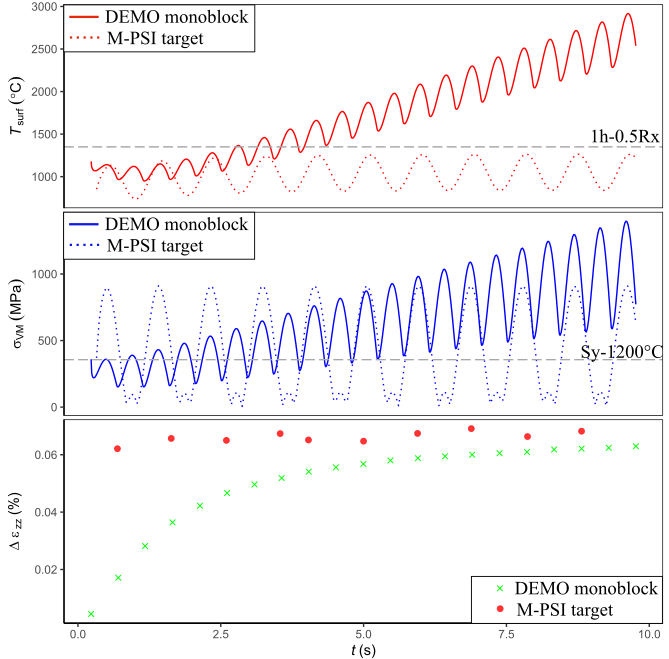}
\caption{Profiles of surface temperature (a), VM stress (b) and true
strain (c) for the monoblock and Magnum-PSI target. 1h-0.5Rx denotes
50\% recrystalisation after 1 hr, and Sy-1200°C denotes yield stress at
1200°C.}
\end{figure}

These FEA results should be interpreted with due care, as experimental
material property data on W above 1200 °C is sparse. There may be
significant variance in the properties used by the simulation.
Additionally, an isothermal bilinear kinematic hardening model was
employed \autocite{zinovevModelingStrainHardening2021}. Errors may also
arise from the assumptions of a constant convective heat transfer
coefficient and perfect thermal contact across dissimilar joints.

\newpage

\subsection{3.2 Magnum-PSI discharges}\label{magnum-psi-discharges}

A total of 12 fatigue testing targets were subjected to sweeping plasma
exposures using Magnum-PSI. Four polished and notched targets received
only sweeping exposures as per the parameters of Table 1. Four other
targets were first exposed to a steady state H implantation phase,
intended to induce blistering and roughening of the plasma-facing
surface. These targets were subsequently exposed to sweeping exposures.
Another four targets received ELM-like loading pre-cracking exposures,
before sweeping exposures.

IR camera measurements of maximum target surface temperature
(\emph{T\textsubscript{surf}}) during a typical Magnum-PSI sweeping
discharge are presented in Fig. 6. To minimise unintentional shock
cracking heating and cooling rates at the start and end of each
discharge were limited to 17 °C s\textsuperscript{-1}. The method
exhibited excellent repeatability and the FEA-determined temperature
range goal of 842 - 1255 °C was consistently achieved. However, a slight
rise in maximum peak temperature was observed during cycling due to heat
accumulation in the target. This was mitigated by implementing a 14 s
dwell phase between every 10 cycles.

OES spectra were monitored continuously and no characteristic emission
lines for W or any other impurities were observed, suggesting that
target sputtering was minimal.

\begin{figure}

{\centering \includegraphics{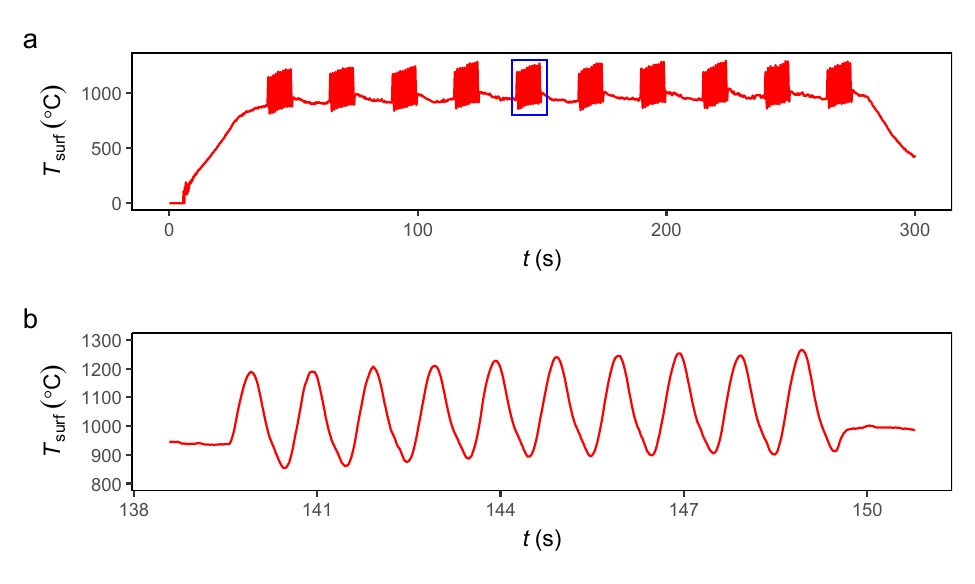} 

}

\caption{Maximum target temperature measured by IR camera during a typical Magnum-PSI sweeping discharge. a) target surface temperature against time for 10 sets of 10 cycles, b) 1 set of 10 (blue box).}\label{fig:unnamed-chunk-2}
\end{figure}

The maxima of 2D Thomson scattering profiles of electron temperature
(\emph{T\textsubscript{e}}) and density (\emph{n\textsubscript{e}}) were
used to determine the maximum heat flux incident on the target
(\emph{q\textsubscript{inc}}). This was taken as the sum of electron
thermal, ion thermal, ion sheath, and surface recombination energies. A
thermalised plasma flow was assumed (\emph{T\textsubscript{e}} =
\emph{T\textsubscript{i}}) with a Mach number of 0.39 and adiabatic
index of 5/3 \autocite{cornelissenErosionEnhancementImpurity2024}. The
surface recombination term included contributions from both atomic (13.6
eV) and molecular (2.2 eV) recombination, and assumed electron energy,
ion energy, and ion particle reflection coefficients of 0.15, 0.4, and
0.6 respectively. Representative exposure data from a single target (300
cycles, no H pre-implantation) are presented in Fig. 7.

\newpage

\begin{figure}

{\centering \includegraphics{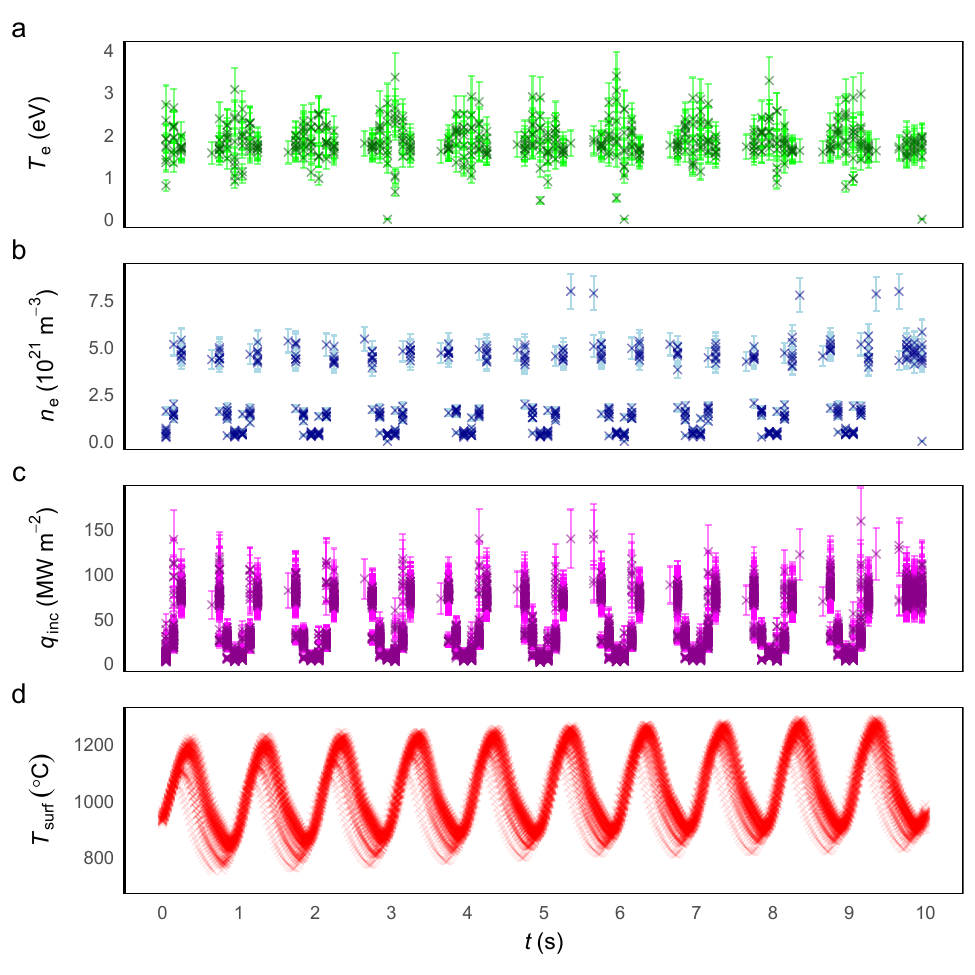} 

}

\caption{Profiles of (a) $T_e$, (b) $n_e$, (c) $q_{inc}$,  and (d) $T_{surf}$ against time for 300 cycles. Thomson and heat flux error bars represent sum of stastical and systematic errors. $T_{surf}$ error bars omitted for clarity.}\label{fig:unnamed-chunk-6}
\end{figure}

The maximum measured heat flux was 159 ± 29 MW m\textsuperscript{-2},
which significantly differs from the 20.25 MW m\textsuperscript{-2}
predicted by the FEA modelling. This discrepancy arises from (i) TS
inaccuracies at high electron density, leading to an overestimate, and
(ii) the expansion and contraction of Magnum-PSI's plasma column during
sweeping, which is neglected by the modelling. After removing outliers
(using 1.5 times the interquartile range), the minimum and maximum FWHM
values as measured by Thomson scattering were 10.6 mm and 17.0 mm
respectively, and the mean FWHM was 13.9 mm.

This variable plasma column width was neglected by the FEA modelling,
hence the peak required heat flux required was underestimated. however,
as the goal temperature range was still achieved, this discrepancy will
not affect the uniaxial strain range imposed on the target or the
validity of the experimental results. Future modelling work will account
for plasma column expansion and contraction.

\newpage

\newpage

\subsection{3.3. Ex-situ crack characterisation and
analysis}\label{ex-situ-crack-characterisation-and-analysis}

\subsubsection{3.3.1. Plasma-facing surfaces prior to sweeping
exposures}\label{plasma-facing-surfaces-prior-to-sweeping-exposures}

Fig. 7. presents SEM images of typical target plasma-facing surfaces
prior to fatigue cracking. Fig. 7a shows an example of crack-free stress
concentration notch prior to any plasma exposure. Fig. 7b shows the
result of the H implantation exposure from Table 1. Fig. 7c shows the
pre-cracked surface of the EDM-cut target series.

\begin{figure}
\centering
\includegraphics[width=1\textwidth,height=\textheight]{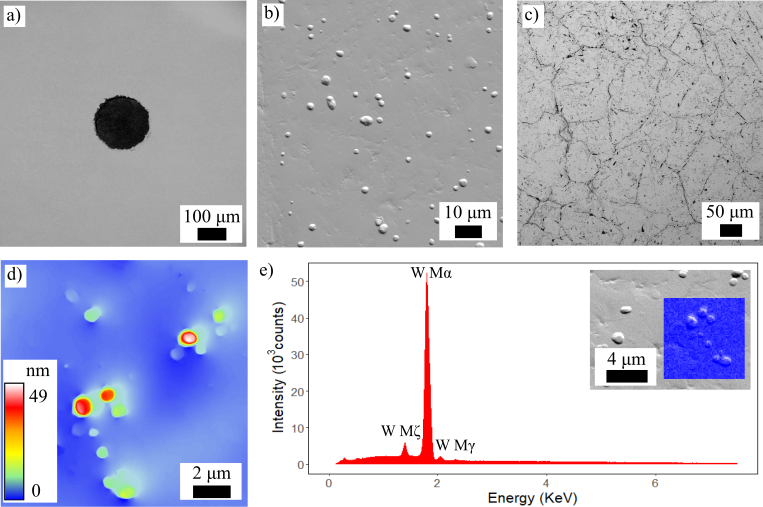}
\caption{SEM characterisation of reference plasma-facing target surfaces
prior to fatigue cracking a) a crack-free notched and polished target
surface pre-implantation. b) A target after H implantation, showing
blistering and roughening of the plasma-facing surface c) a typical
pre-cracked EDM-cut surface d) BSE quad surface height reconstruction of
blisters e) EDX spectra of a blistered region (inset in blue).}
\end{figure}

After H implantation as per Table 1 settings targets exhibited surface
roughening and the formation of ellipical surface blisters (Fig. 8b).
Quantitative SEM image analysis of blistered regions found the mean
blister Feret diameter to be 2.02 µm with a standard deviation of 0.725
µm (n=26). BSE quad height reconstruction was used to estimate blister
heights, which was between 15 and 49 nm (Fig. 9d). These blister
dimensions are consistent with those observed in previous Magnum-PSI
experiments \autocite{chenGrowthMechanismSubsurface2020}. To verify that
theses features were blisters and not surface contamination
(e.g.~similarly-sized SiO\textsubscript{2} polishing media) EDX of a
blistered region was performed (Fig. 9e). This was confirmed by
exclusive observation of W's characteristic Mζ, Mα and Mγ peaks.

The surface of the EDM cut target series (Fig. 8c) exhibited a network
of fine micro-cracks 100-300 µm in length. Several shallow surface pits
were observed approx. 54-77 µm diameter where flakes of W are assumed to
have broken away from the surface. The surface finish of these targets
closely resembled the technical surface finish of ITER-representative
monoblock mockup chains previously exposed in Magnum-PSI
\autocite{morganITERMonoblockPerformance2020}.

\newpage

\subsubsection{3.3.2 Polished targets with and without H
pre-implantation exposed to strikepoint
sweeping}\label{polished-targets-with-and-without-h-pre-implantation-exposed-to-strikepoint-sweeping}

All four targets without H pre-implantation (hereafter referred to `no
prior H') were found to exhibit large mm-scale surface cracks after
sweeping exposures (\emph{N} = 150, 300, 450, 600). In each case large
mm-scale fatigue cracks originated from the the central stress
concentration notch and propagated radially towards the target edge.
This relatively fast crack propagation suggests that non-linear rapid
crack growth has occurred. Multiple instances of crack branching were
observed on all cracked targets, however there appeared to be no
observable pattern in branching behaviour.

Stitched backscattered electron (BSE) micrographs of targets with and
without prior H implantation are shown in Fig. 9. The greater
electron-matter interaction depth of BSE mode relative to secondary
electron (SE) mode was exploited to mask surface blistering (which would
interfere with image thresholding) while maintaining visability of the
relatively deeper fatigue cracks. Target \emph{N} = 300 of the
\emph{prior H} series, and targets \emph{N} = 300, 450 of the \emph{no
prior H} series are omitted from Fig. 9 for brevity, but exhibited
similar cracking behaviour.

\begin{figure}
\centering
\includegraphics[width=1\textwidth,height=\textheight]{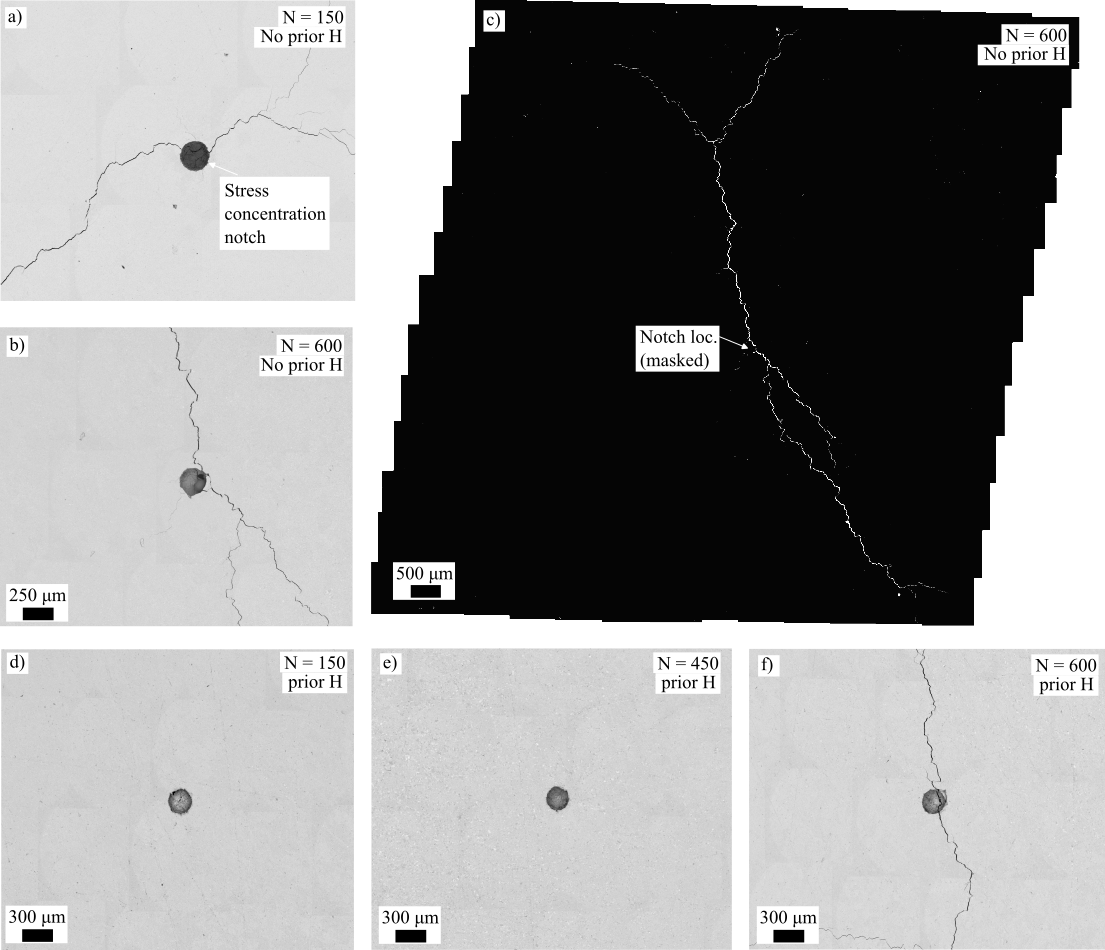}
\caption{Montage of stitched BSE images showing macroscopic fatigue
cracking in the No prior H (a, b) and prior H (d, e, f) series. Fig. 9c
shows a full-scale thresholded stitched BSE image prior to quantitative
image analysis. Square features of approx. 1 mm size are
stitching/contrast artifacts.}
\end{figure}

\newpage

Of the four targets which received prior H implantation (as per Table
2), only the \emph{N} = 600 cycle target exhibited macroscopic mm-scale
cracking. Targets with \emph{N} = 150, 300 and 450 exhibited only
micro-scale cracking local to within 100 µm of the target centre. Fig.
10 shows SE mode micrographs of the \emph{N} = 600 target of the prior H
implantation series.

\begin{figure}
\centering
\includegraphics[width=1\textwidth,height=\textheight]{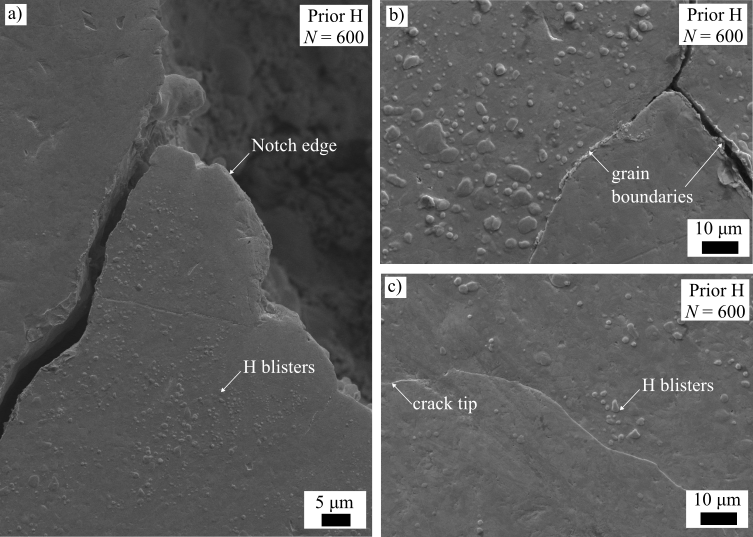}
\caption{SE micrographs of a cracked and blistered target. a) crack
initiation at the notch b) blistering and inter-granular cracking at a
branch c) detail of propagation behaviour at the crack tip.}
\end{figure}

Fig. 9a shows the edge of a notch with the initiation site of a mm-scale
fatigue crack with a maximum width of approximately 4 µm. Fig. 9b
presents a typical region along the crack path showing both blistering
and crack propagation. Intergranular cracking may be inferred from the
absence of blistering on the lower region of the image, and prior work
which reports that blistering occurs on preferred crystallographic
orientations due to the channelling effect
\autocite{dubinkoSubSurfaceMicrostructureSingle2017}.

The tip of the same fatigue crack is shown in Fig. 9c. If stress
concentrations at the edges of blisters were altering crack propagation
behaviour or causing the formation of a micro-crack network, the crack
tip could be expected to propagate via blisters. That this is not
observed suggests that the presence of H blisters does not have a
significant effect on fatigue crack propagation. This is mostly likely
due to their relatively small scale resulting in insufficient stress
concentration to affect fatigue crack propagation. It is therefore also
unlikely that nano-scale W fuzz will alter cracking behaviour, however
this should be confirmed experimentally by future work.

\newpage

The results of quantitative SEM image analysis via thresholding and
ridge detection are presented in Fig. 11. As could be expected from the
Paris-Erdogan equation, \emph{L} and \emph{ρ} both exhibit a power law
relationship with respect to \emph{N}. The main source of error is
likely different degrees of polishing between targets, which results in
noise on the thresholding.

\begin{figure}

{\centering \includegraphics{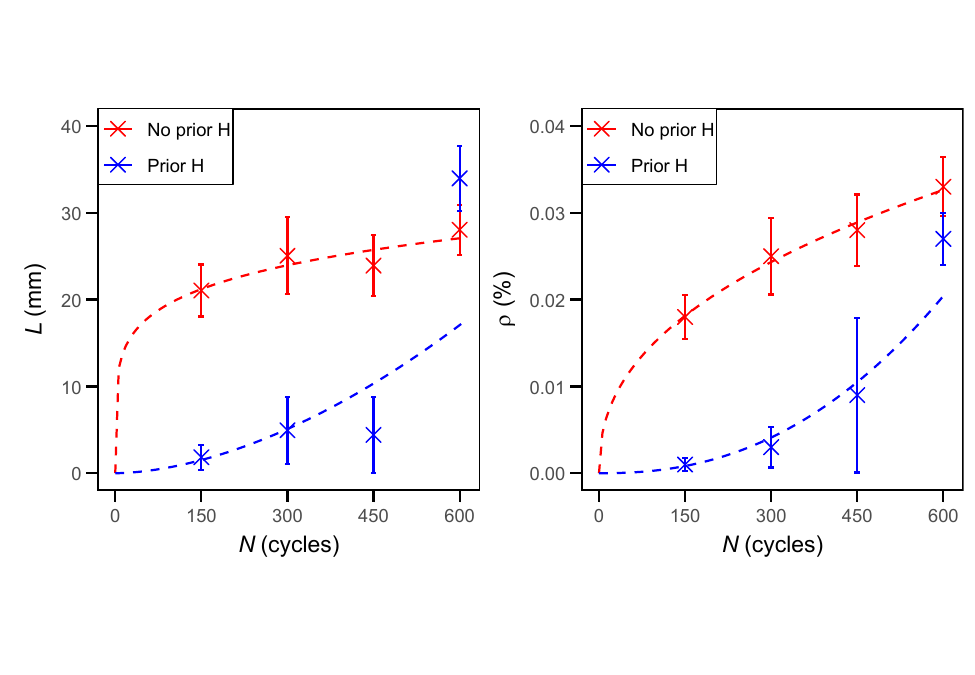} 

}

\caption{Total crack length (L) and crack density (ρ) for targets with and without prior H implantation. Errors bars represent the sum of stitching repeatability and thresholding process errors. Parameters for the power law fits are given in Table 3.}\label{fig:unnamed-chunk-8}
\end{figure}

The prior H series exhibits a significantly slower rate of crack growth
than the No prior H series, which suggests that the H implantation
impedes fatigue cracking. At exposure temperatures of 850-1250 °C it's
likely that all of the implanted H will have outgassed from the
super-saturated surface
\autocite{alimovTemperatureDependenceSurface2012}. However, the
micro-scale voids and dislocations induced by H implantation may remain.
Published TEM micrographs of a W surface exposed to identical plasma
conditions revealed a greatly increased dislocation density local to
blisters and voids \autocite{chenGrowthMechanismSubsurface2020}. It is
theorised that this increase in dislocation density at the plasma-facing
surface may induce a \emph{case hardenening} effect, whereby dislocation
entanglement (self-pinning) inhibits glide and PSB formation, thereby
delaying fatigue crack initiation.

\begin{table}[h!]
\caption{Power law fitting parameters for the fits of Fig. 11.}
\begin{tabular}{|c|c|c|c|c|c|c|}
\hline
Series & \multicolumn{3}{c|}{Length} & \multicolumn{3}{c|}{Density} \\
\cline{2-7}
 & a & b & R\textsuperscript{2} & a & b & R\textsuperscript{2}\\
\hline
No prior H & 8.78 & 0.176 & 0.796 & 2.17\times 10\textsuperscript{-3} & 0.424 & 0.99\\
Prior H & 2.43\times 10\textsuperscript{-4} & 1.745 & 0.728 & 7.68\times 10\textsuperscript{-9} & 2.312 & 0.96\\
\hline
\end{tabular}
\end{table}

\newpage

\subsubsection{3.3.3 Targets with pre-cracked (ELM-like) surfaces
exposed to strikepoint
sweeping}\label{targets-with-pre-cracked-elm-like-surfaces-exposed-to-strikepoint-sweeping}

The combined effects of type I or II ELM micro-cracking and strikepoint
sweeping were also investigated. This employed two different methods to
induce global and local ELM-like pre-cracking of the plasma-facing
surface. The global method employed electo-discharge machining to
introduce a degree of recrystalisation and a network of fine
micro-cracks across the entire surface. The local method used a Nd:YAG
laser to induce cracks in a 2.5 mm diameter region, as per
\autocite{detemmermanHighHeatFlux2013}. Both sets of pre-cracked targets
were subsequently exposed to sweeping exposures as per the parameters of
Table 1. Figs. 11a, 11b and 11c present SEM micrographs of the globally
pre-cracked series after 20, 100 and 1000 cycles of 850-1250 °C
respectively.

\begin{figure}
\centering
\includegraphics[width=1\textwidth,height=\textheight]{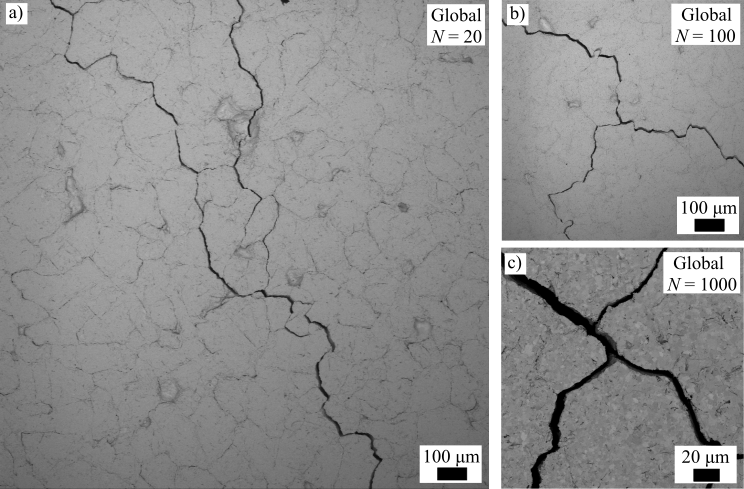}
\caption{The effects of strikepoint sweeping on EDM-induced ELM-like
pre-cracked surfaces after a) 20 cycles, b) 100 cycles and c) 1000
cycles. The polishing effect observed in Fig. 12c is due to
unintentional sputtering by impurities from Magnum-PSI's plasma source.}
\end{figure}

After just 20 cycles large mm-scale cracks had emerged from the centre
of the target (fig.~11a). These dominant cracks preferentially
propagated along the prior microcrack network radially from the centre.
This is in contrast to the inter-granular cracking exhibited by polished
targets (section 3.3.2), which suggests that ELM micro-crack coalescence
may contribute to the accelerated propagation of fatigue cracks. Similar
cracking behaviour was observed for the 100 and 1000 cycle targets, in
addition to crack widening.

Fig. 12a shows a the surface of a polished and notched target after
exposure to 1000 ELM-like laser pulses (parameters as per table 2),
followed by 100 cycles of strikepoint sweeping at 850-1250 °C. Laser
loading was found to have induced an ellipsoidal region of fine
micro-cracks 100-500 µm in length. This region had a Feret diameter of
2.5 mm and was located 2.1 mm away from the stress concentration notch
due to mis-alignment of the laser. Emerging from this region were
several large fatigue cracks 3-6 mm in length.

\newpage

Several examples of cracking-related phenomena of concern were observed
from the localised pre-cracked target. Fig. 13b shows a loosened W flake
with a Feret diameter of 33 µm liberated by fatigue cracking. In Fig.
13c a large surface pit with a Feret diameter of 83 µm can be observed,
suggesting that a relatively large flake of W may have been liberated
from the surface. The co-location of the pit with the path of a fatigue
crack suggests that cracking is responsible for the flaking and material
ejection, which is further confirmed by the presence of fatigue
striations on the surface of the pit (Fig. 13d).

Fig. 13e shows a solidified W droplet with a Feret diameter of 48 µm
located in a 70-100 µm dia. pit. Cracking has resulted in the thermal
isolation of a region of W from the bulk, giving rise to localised
heating and melting, followed by droplet formation. The presence of
striations (detail red box) again indicates fatigue as the root cause,
and visible recrystalisation of the nearby connective material evidences
localised heating.

\begin{figure}
\centering
\includegraphics[width=1\textwidth,height=\textheight]{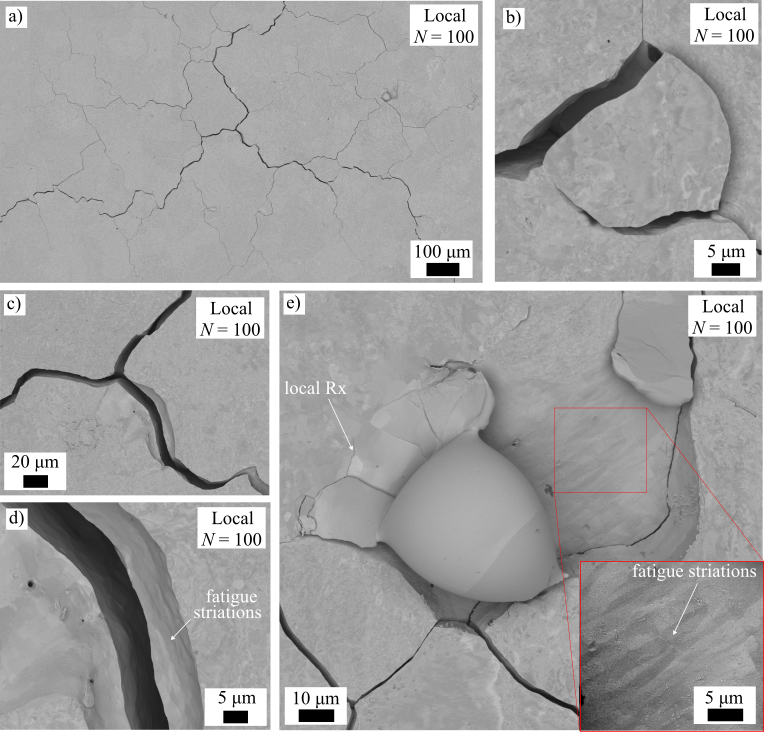}
\caption{Combined effects of localised ELM-like cracking and strikepoint
sweeping a) the ELM-cracked region b) A loosened W flake c, d) A pit
arising from flake liberation e) A solidified W droplet arising from
fatigue}
\end{figure}

\newpage

\section{4. Conclusions}\label{conclusions}

A campaign of novel uniaxial fatigue testing experiments has been
conducted using the Magnum-PSI linear plasma device. A
DEMO-representative experiment was designed using the results of
supporting finite element analyses, which explored the thermomechanical
response of a DEMO monoblock to a reattachment thermal load of 45 MW
m\textsuperscript{-2}, swept along divertor targets at 1 Hz and 100 mm.
ITER-grade W targets were exposed to 150, 300, 450 and 600 thermal
cycles designed to induce the corresponding strain range (Δε). This new
experimental method was used to explore the combined effects of
strikepoint sweeping with prior H implantation, and global and localised
ELM pre-cracking.

Several key findings are reported. Prior H implantation at
\(\Phi = 2.2 \times 10^{26}\) m\textsuperscript{-2} and
\(T_{surf} = 300\) °C was found to significantly delay the subsequent
onset of thermal fatigue cracking in W. H-implanted tungsten specimens
exhibited delayed fatigue crack initiation (450-600 cycles) compared to
non-implanted tungsten, which cracked after fewer than 150 cycles. This
was contrary to existing literature on ELM-like thermal shock
experiments, which found hydrogen exposure enhanced W's susceptibility
to ELM-induced cracking \autocite{wirtzImpactCombinedHydrogen2015}. The
contrasting findings of this study likely arise from the
dislocation-mediated nature of low-cycle fatigue crack initiation. Local
plastic deformation of the H super-saturated surface region induces
dislocations, giving rise to hydrogen induced dislocation pinning. This
results in a case hardening effect, whereby increased dislocation
density and entanglement near the plasma-facing surface inhibits
dislocation glide and persistent slip band formation, thereby delaying
fatigue crack initiation. This hypothesis is supported by TEM analysis
(by others) which confirms that H exposure significantly increases
dislocation density local to blisters and voids
\autocite{chenGrowthMechanismSubsurface2020}. Fatigue crack tips were
observed to propagate independently of surface blisters, suggesting that
local stress fields induced by blister edges were not sufficiently large
enough to alter crack propagation behaviour. From this, it can be
inferred that nano-scale He-induced W fuzz is also unlikely to
significantly affect the fatigue life of DEMO's divertor, but this
should be confirmed experimentally.

A limited exploration of the combined effects of strikepoint sweeping
and ELM-like pre-cracking was undertaken, and a synergistic interaction
was revealed. Strikepoint sweeping over an ELM-cracked surface was found
to result in the coalescence of ELM-induced micro-cracks into larger
dominant fatigue cracks, which rapidly propagated at the
multi-millimeter scale along the pre-existing microcrack network.
Flaking, localised melting and formation of W droplets up to 30 µm in
diameter were also observed. A proposed mechanism for this is the
thermal isolation of a region of W by fatigue cracking, which then melts
despite the bulk surface temperature being well below W's melting point
of 3422 °C \autocite{zhuSituLeadingEdgeInducedDamages2022}. The
formation and liberation of such droplets poses a significant risk of
tungsten transport to the scrape-off layer and potential migration to
the plasma core, which could trigger major plasma disruptions. Droplet
ejection from a damaged W surface has been experimentally observed in
both TEXTOR and EAST discharges
\autocite{coenenMeltLayerEjectionMaterial2011,zhuSituLeadingEdgeInducedDamages2022}.

In summary, the following conclusions may be drawn from this work. (i)
Stress concentrations arising from 1-10 µm blistering does not appear to
significantly alter the propagation behaviour of fatigue cracks. (ii)
However, H implantation appears to significantly improve W's resistance
to fatigue cracking, likely due to a case hardening effect arising from
an increased dislocation density in the H supersaturated surface region
which remains after high temperature H outgassing. (iii) The combination
of ELM micro-cracking and strikepoint sweeping appears to result in the
very rapid (\emph{N} \textless20) micro-crack coalescence and
propagation of mm-scale fatigue cracks, which may lead to flaking,
localised melting and formation of W droplets.

This work has demonstrated the feasibility of using a linear plasma
device to perform DEMO-representative uniaxial fatigue testing of
plasma-facing materials. It provides valuable insights into the
contributing factors of fatigue cracking of plasma-facing surfaces, and
the possible modes of failure of DEMO's divertor monoblocks arising from
cyclic thermomechanical loading. There remains significant scope for
future work, in both refining the experimental method, expanding the
scope to novel plasma-facing material concepts, and further exploration
of proposed strikepoint sweeping regimes for DEMO. There remains much
scope future work on this topic, some possible avenues for which are
discussed in Section 5.

\section{5. Scope for future work}\label{scope-for-future-work}

Future work on this topic could explore the effects of other
plasma-material interactions, such as He-induced W fuzz, impurity
sputtering (i.e.~mixed composition plasmas), and the fatigue behaviour
of redeposited W layers. The synergistic effects of neutron irradiation,
high-temperature creep, and recrystallization should also be
investigated. Testing of advanced divertor material concepts such as
tungsten fiber-reinforced tungsten composites under simultaneous plasma
and fatigue loading conditions would also be valuable for evaluating
their fatigue resistance
\autocite{neuMaterialComponentDevelopments2023}. The effects on liquid
metal capillary porous structure (CPS) alternative divertor concepts
could also be explored.

Scope remains for refinement of both the experimental method and
supporting modelling. Implementing a linear ramp in addition to the
modulating plasma would more closely mimic the thermomechanical loading
profile of a real tokamak reattachement event. The present approach
over-estimates the strain range, which may lead to unnecessary design
conservatism. Direct measurement of target strain using a time of flight
(ToF) or triangulation laser rangefinding technique would eliminate any
modelling-induced errors and improve confidence in the experimental
results. Additionally, higher model fidelity may be achieved by
accounting for radiation embrittlement, and extending the simulation
time to multiple reattachment events to account for time-dependent
strain hardening and creep.

Perhaps most importantly, a diagnostic for in-situ micro-crack detection
should be developed which facilitates precise determination of the
number of cycles to failure. This diagnostic would enable more rapid and
material-efficient testing, and could utilise infrared thermography
techniques such as lock-in or flying spot thermography, an embedded eddy
current or ultrasonic system, or an ultra-long working distance light
optical microscope
\autocite{yuSignalDistributionBasedCrackDetection2024}. This diagnostic
is a pre-requisite to development of a comprehensive database of fatigue
behaviour (ε-N curves), and would provide the necessary experimental
data for constituitive modelling of PMI effects on fatigue crack
initiation and propagation.

\section{6. Acknowledgements}\label{acknowledgements}

The Magnum-PSI operations team and DIFFER's Plasma-Material Interactions
group are thanked for their support, especially Cas Robben and Jos
Scholte. Magnum-PSI is funded by the Netherlands Organisation for
Scientific Research (NWO) and Euratom. This work has been carried out
within the framework of the EUROfusion Consortium, funded by the
European Union via the Euratom Research and Training Programme (Grant
\#\#\#.\#\#\#\#.\#\#\#.). Views and opinions expressed are however those
of the author(s) only and do not necessarily reflect those of the
European Union or the European Commission.

\section{7. Declarations and
statements}\label{declarations-and-statements}

\subsection{7.1 CrediT authorship contribution
statement}\label{credit-authorship-contribution-statement}

\textbf{J. Hargreaves}: Conceptualisation, methodology, investigation,
data curation, formal analysis, visualisation, Writing -- original
draft, project administration. \textbf{J.H You}: Conceptualisation,
supervision, validation, writing -- review \& editing. \textbf{F.
Maviglia}: Conceptualisation, supervision, validation, writing -- review
\& editing. \textbf{J. Vernimmen}: Methodology, data curation, writing
-- review \& editing. \textbf{J. Scholten}: Methodology, writing --
review \& editing. \textbf{T. W. Morgan}: Funding acquisition,
supervision, conceptualisation, project administration, writing --
review \& editing

\subsection{7.2 Declaration of competing
interest}\label{declaration-of-competing-interest}

The authors declare that they have no known competing financial
interests or personal relationships that could have appeared to
influence the work reported in this paper.

\subsection{7.3 Declaration of Generative AI and AI-assisted
technologies
use}\label{declaration-of-generative-ai-and-ai-assisted-technologies-use}

During the preparation of this work the author(s) used Anthropic Claude
3.7 Sonnet as a coding and data analysis assistant. This work contains
no AI-generated text or images. After using this tool/service, the
author(s) reviewed and edited the content as needed and take(s) full
responsibility for the content of the publication.

\printbibliography[title=8. Bibliography]

\end{document}